\documentclass[12pt]{article}
\usepackage{amsmath}
\usepackage{graphics,graphicx,xcolor}
\def\be{\begin{equation}}
\def\ee{\end{equation}}
\def\arr{\begin{array}{rll}}
\def\ea{\end{array}}
\def\bea{\begin{eqnarray}}
\def\eea{\end{eqnarray}}

\def\N2{$N{=}2$}

\def\>{\rangle}
\def\<{\langle}
\def\+{\dagger}
\def\={\ =\ }

\begin{document}
\renewcommand{\thefootnote}{\fnsymbol{footnote}}
\begin{titlepage}
\setcounter{page}{0}
\vskip 1cm
\begin{center}
{\LARGE\bf Dynamical realizations of the Lifshitz group}\\
\vskip 1.2cm
$
\textrm{\Large Anton Galajinsky\ }
$
\vskip 0.7cm
{\it
Tomsk State University of Control Systems and Radioelectronics, 634050 Tomsk, Russia} \\

\vskip 0.2cm
{e-mail: a.galajinsky@tusur.ru}
\vskip 0.5cm

\end{center}
\vskip 1cm
\begin{abstract} \noindent
Dynamical realizations of the Lifshitz group are studied within the group--theoretic framework. A generalization of the $1d$ conformal mechanics is constructed, which involves an arbitrary dynamical exponent $z$. A similar generalization of the Ermakov--Milne--Pinney equation is proposed. Invariant derivative and field combinations are introduced, which enable one to construct a plethora of dynamical systems enjoying the Lifshitz symmetry. A metric of the Lorentzian signature in $(d+2)$--dimensional spacetime and the energy--momentum tensor are constructed, which lead to the generalized Ermakov--Milne--Pinney equation upon imposing the Einstein equations. The method of nonlinear realizations is used for building Lorentzian metrics with the Lifshitz isometry group. In particular, a $(2d+2)$--dimensional  metric is constructed, which enjoys an extra invariance under the Galilei boosts.
\end{abstract}


\end{titlepage}

\renewcommand{\thefootnote}{\arabic{footnote}}
\setcounter{footnote}0

\noindent
{\bf 1. Introduction}\\

\noindent
The non--relativistic version of the AdS/CFT--correspondence \cite{DTS,BM} extends the holographic dictionary to encompass strongly coupled condensed matter systems. It relies upon finite--dimensional conformal extensions of the Galilei algebra, the most general of which is the $\ell$--conformal Galilei algebra \cite{Henkel,NOR}. The latter builds upon
generators of time translation, dilatation, and special conformal transformation, which form $so(2,1)$ subalgebra, spatial rotations, as well as a chain of vector generators describing spatial translations, Galilei boosts, and constant accelerations.

If the special conformal transformation is discarded, the generators of Galilei boosts and constant accelerations can be omitted as well. Furthermore, the commutator of temporal translation and dilatation can be deformed to include an arbitrary constant $z$ known as the dynamical exponent, giving rise to the Lifshitz algebra (see e.g. \cite{MT}). Initiated in \cite{KLM}, the Lifshitz holography attracted considerable
attention\footnote{Literature on the subject is overwhelmingly large. For a review of the field prior to 2016 see \cite{MT}. Interesting accounts of the Lifshitz geometries can be found in \cite{BHR,HW}.} and active research in this direction continues to date.

The goal of this work is to explore dynamical realizations of the Lifshitz group within the group--theoretic framework \cite{CWZ}.

In the next section, a real Lie algebra formed by the generators of translation and dilatation in one--dimensional space is considered and a generalization of the conformal mechanics \cite{DFF} is constructed which involves an arbitrary dynamical exponent $z$. The general solution to the corresponding equation of motion is given in terms of the hypergeometric function \cite{WW}. A transformation of the temporal variable and the dilaton field is given which links the model to the system with $z=1$ in \cite{DFF}.

In Sect. 3, the $1d$ conformal mechanics in a harmonic trap is discussed. A generalization involving arbitrary dynamical exponent $z$ is proposed which is further used to define a generalized Ermakov--Milne--Pinney equation.

Sect. 4 is focused on dynamical realizations of the Lifshitz group in mechanics. In particular, the method of nonlinear realizations \cite{CWZ} is used to determine the invariant derivative and field combinations, which are the building blocks for constructing dynamical systems. In general, such models describe a particle moving in a $d$--dimensional space, which is driven by the conformal mode introduced in Sect. 2. In this setting, the latter acts as a kind of a cosmic scale factor.

Sect. 5 adds to the recent studies in \cite{G,CGGH}, which incorporated a cosmic scale factor within Eisenhart's approach \cite{E} (see also \cite{DBKP,DGH}) to geometrization of classical mechanics. A metric of the Lorentzian signature in $(d+2)$--dimensional spacetime and the energy--momentum tensor are constructed, which result in the generalized Ermakov--Milne--Pinney equation upon imposing the Einstein equations.
The corresponding null geodesics describe a variant of the Lifshitz oscillator driven by the conformal mode.

In Sect. 6, the group--theoretic construction is applied to build Lorentzian metrics possessing the Lifshitz isometry group. In particular, we reproduce the $(d+2)$--dimensional metric in \cite{KLM} as well as construct its $(2d+2)$--dimensional extension enjoying an extra invariance under the Galilei boosts.

In Appendix A, symmetries of the $1d$ conformal mechanics in a harmonic trap are discussed. Appendix B contains a group--theoretic analysis of the Lifshitz algebra extended by the generator of Galilei boosts.

Throughout the paper, summation over repeated indices is understood.

\vspace{0.5cm}

\noindent
{\bf 2. The conformal mode}\\

\noindent
First, it is worth reminding how the conventional $1d$ conformal mechanics is introduced in \cite{DFF}.
Consider the conformal transformation in one dimension
\be\label{sl2}
t'=\frac{\alpha t+\beta}{\gamma t+\delta}, \qquad \alpha \delta-\beta \gamma=1,
\ee
a primary field $\rho(t)$ of the conformal weight $\frac 12$
\be\label{trR}
\rho' (t')={\left(\dot{t'}\right)}^{\frac 12} \rho(t),
\ee
where the dot designates the derivative with respect to $t$,
and the action functional \cite{DFF}
\be\label{DFF}
S=\frac 12 \int dt \left(\dot\rho^2-\frac{\gamma^2}{\rho^{2}} \right),
\ee
with a constant $\gamma$. Being applied to (\ref{DFF}), the transformations (\ref{sl2}) and (\ref{trR}) yield
\be
S'= S+\frac 14 \int dt \left( {\left( \rho^2 \frac{\ddot{t'}}{\dot{t'}}\right)}^{\dot{}} -\rho^2 \left(\frac{\dddot{t'}}{\dot t'}-\frac 32 {\left(\frac{\ddot t'}{\dot t'} \right)}^2 \right) \right).
\ee
The last term involves the Schwarzian derivative $\frac{\dddot{t'}}{\dot t'}-\frac 32 {\left(\frac{\ddot t'}{\dot t'} \right)}^2$, which is known to vanish for the $SL(2,R)$ transformation in (\ref{sl2}), while the second term is a total derivative.
Thus, the action (\ref{DFF}) describes a self--interacting $1d$ conformal field theory, $\gamma$ being a coupling constant.  Alternatively, the system can be regarded as a particle moving on a real line parameterized by the coordinate $\rho$ in the external field potential $U(\rho)=\frac{\gamma^2}{\rho^{2}}$. In the latter interpretation, $\gamma$ links to the strength of the external force.

As was demonstrated in \cite{IKL}, the model (\ref{DFF}) can be obtained by applying the conventional group--theoretic construction \cite{CWZ} to the Lie algebra $sl(2,R)\simeq so(2,1)$.
Given the structure relations of $sl(2,R)$
\be\label{sl2r}
[H,D]={\rm i} H, \qquad [H,K]=2 {\rm i} D, \qquad [D,K]={\rm i} K,
\ee
where $H$, $D$, $K$ generate translation, dilatation, and special conformal transformation, respectively,
one considers the group--theoretic element
\be\label{g}
g=e^{{\rm i}tH} e^{{\rm i} s(t) K} e^{{\rm i} u(t) D},
\ee
where $t$ is a temporal variable and $s(t)$, $u(t)$ are the Goldstone fields. Using
the Baker--Campbell--Hausdorff formula
\be\label{ser}
e^{iA}~ T~ e^{-iA}=T+\sum_{n=1}^\infty\frac{i^n}{n!}
\underbrace{[A,[A, \dots [A,T] \dots]]}_{n~\rm times},
\ee
one computes the Maurer--Cartan one--forms
\be
g^{-1} d g={\rm i} \omega_H H+{\rm i} \omega_K K+{\rm i} \omega_D D,
\ee
where
\be\label{inv}
\omega_H=e^{-u} dt, \qquad \omega_K=e^u \left(\dot s+s^2\right)dt, \qquad \omega_D=\left(\dot u-2 s \right) dt,
\ee
which hold invariant under the $SO(2,1)$--transformation $g'= e^{{\rm i}\alpha H} e^{{\rm i} \sigma  K} e^{{\rm i} \beta D} \cdot g$ parameterized by real numbers $\alpha$, $\beta$, and $\sigma$.

Introducing a new field $\rho=e^{\frac{u}{2}}$ and imposing the $SO(2,1)$--invariant constraints \cite{IKL}
\be\label{const}
\omega_D=0, \qquad \omega_K=\gamma^2 \omega_H,
\ee
where $\gamma$ is interpreted as a coupling constant, one can use the first condition in (\ref{const}) to eliminate $s=\frac{\dot\rho}{\rho}$ from the consideration, while the second restriction reproduces the conformal mechanics equation of motion \cite{DFF}
\be\label{CM}
\ddot\rho=\frac{\gamma^2}{\rho^3}.
\ee

Before we turn to a generalization of (\ref{CM}) involving an arbitrary dynamical exponent $z$, it proves instructive to obtain (\ref{CM}) without invoking the generator of special conformal transformation $K$ as the latter does not belong to the Lifshitz algebra. Note also that, as far as (\ref{CM}) is concerned, a constant of the motion associated with the special conformal transformation is functionally dependent on other integrals of motion \cite{DFF} and, hence, can be discarded.

Setting $s=0$ in (\ref{inv}) and specifying to a subalgebra formed by $H$ and $D$, one gets the invariant derivative $\mathcal{D}=e^{u} \frac{d}{dt}$ and the invariant field $\mathcal{D} u$. Using them to construct the equation of motion
\be\label{Neq}
\mathcal{D}^2 u+h  {\left(\mathcal{D} u\right)}^2=2 \gamma^2,
\ee
where $h$ and $\gamma$ are real constants, one can fix $h$ from the requirement that (\ref{Neq}) takes the conventional conservative mechanics form $\ddot\rho=-\frac{\partial U(\rho)}{\partial\rho}$ after introducing $\rho=e^{\frac{u}{2}}$. This yields $h=-\frac 12$ and reduces (\ref{Neq}) to (\ref{CM}).

We are now in a position to formulate a generalization of the conformal mechanics (\ref{CM}) which involves an arbitrary dynamical exponent $z$. Let us modify the first commutator in (\ref{sl2r}) in accord with the Lifshitz algebra
\be\label{HDz}
[H,D]={\rm i} z  H.
\ee
Introducing the group--theoretic element
\be\label{g}
g=e^{{\rm i}tH} e^{{\rm i} u(t) D},
\ee
and repeating the steps above, one finds the invariant derivative $\mathcal{D}=e^{z u} \frac{d}{dt}$ and the invariant field $\mathcal{D} u$, while Eq. (\ref{Neq}) yields\footnote{Because in general $4z-1$ is an arbitrary real number, one has to assume that $\rho$ is dimensionless, while $[\gamma]=[t^{-1}]$.}
\be\label{LM}
\ddot\rho=\frac{(2 z-1) \gamma^2}{\rho^{4z-1}},
\ee
after setting $\rho=e^{\frac{u}{2}}$, $h=\frac{1-2z}{2}$ (the latter condition removes the ${\dot\rho}^2$--term from the equation of motion) and rescaling the coupling constant $\frac{\gamma^2}{2z-1} \to \gamma^2$.

Symmetries of (\ref{LM}) are obtained from
$g'= e^{{\rm i}\alpha H} e^{{\rm i} \beta D} \cdot g$, which gives
\bea\label{TR}
&&
t'=t+\alpha, \qquad \rho'(t')=\rho(t);
\nonumber\\[2pt]
&&
t'=e^{\beta z} t, \qquad  ~ \rho'(t')=e^{\frac{\beta}{2}}\rho(t),
\eea
where $\alpha$ and $\beta$ are real finite parameters.
Note that (\ref{LM}) reduces to (\ref{CM}) at $z=1$.

The general solution to (\ref{LM}) is found by integrating a first order differential equation which follows from the expression for the conserved energy
\be\label{GS}
E=\frac 12 \left(\dot\rho^2+\frac{\gamma^2}{\rho^{4z-2}}\right) \quad \Rightarrow \quad \pm \left(t-t_0 \right)= \frac{ {}_2 F_1 \left(\frac 12,\frac{1}{2-4z};1+\frac{1}{2-4z};\frac{\gamma^2}{2 E \rho^{4z-2}} \right) \cdot \rho}{\sqrt{2E}},
\ee
where $t_0$ is a constant of integration and ${}_2 F_1(a,b;c;x)$ is the hypergeometric function \cite{WW}.\footnote{Because the hypergeometric series ${}_2 F_1(a,b;c;x)$ is ill defined for $c=-n$, $n$ being a natural number, a decreasing sequence
$z_n=\frac{3+2n}{4(1+n)}$, which starts at $\frac{5}{8}$ and converges to $\frac{1}{2}$,
will be excluded from the consideration. Note also that the hypergeometric series, which specifies ${}_2 F_1(a,b;c;x)$, converges for $|x|<1$ only \cite{WW}. The domain of the function in (\ref{GS}) is consistent with the convergence condition.}
When obtaining (\ref{GS}), the following identity
\be
{}_2 F_1 \left(\frac 12,\frac{1}{2-4z};1+\frac{1}{2-4z};x \right)+(2-4z) x \frac{d}{dx} \left( {}_2 F_1 \left(\frac 12,\frac{1}{2-4z};1+\frac{1}{2-4z};x \right) \right)=\frac{1}{\sqrt{1-x}}
\nonumber
\ee
proved helpful. Fig. 1 displays the graph of $t=\frac{ {}_2 F_1 \left(\frac 12,\frac{1}{2-4z};1+\frac{1}{2-4z};\frac{\gamma^2}{2 E \rho^{4z-2}} \right) \cdot \rho}{\sqrt{2E}}$ for $E=\frac 12$, $\gamma=1$, $z=0.8$ (bottom), $z=1$ (middle), $z=1.2$ (top).

\begin{figure}[ht]
\begin{center}
\resizebox{0.5\textwidth}{!}{%
\includegraphics{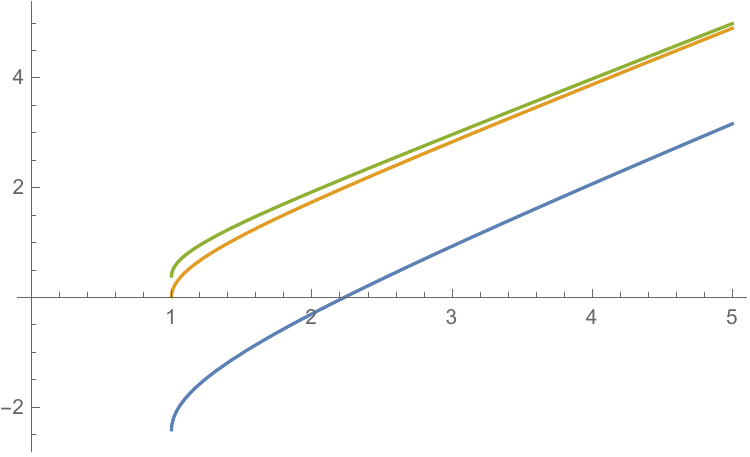}}\vskip-4mm
\caption{\small The graph of $t=\frac{ {}_2 F_1 \left(\frac 12,\frac{1}{2-4z};1+\frac{1}{2-4z};\frac{\gamma^2}{2 E \rho^{4z-2}} \right) \cdot \rho}{\sqrt{2E}}$ for $E=\frac 12$, $\gamma=1$, $z=0.8$ (bottom), $z=1$ (middle), $z=1.2$ (top).}
\label{fig3}
\end{center}
\end{figure}

In order to keep the parallel with the conformal mechanics, in what follows we assume $z>\frac 12$
such that (\ref{LM}) describes scattering off the center. In particular, the numerical value of $z$ can be used to engineer a distance from the center $\rho=0$ to the turning point at
\be\label{TP}
\rho_0 (z)={\left( \frac{\gamma^2}{2E} \right)}^{\frac{1}{4z-2}} \quad \Rightarrow \quad \rho_0 (z)={\rho_0 (1)}^{\frac{1}{2z-1}}.
\ee
Fig. 2 displays the graph of $U(\rho)=\frac{\gamma^2}{2 \rho^{4z-2}}$ for $\gamma=1$, $z=0.8$ (top), $z=1$ (middle), $z=1.2$ (bottom).

\begin{figure}[ht]
\begin{center}
\resizebox{0.5\textwidth}{!}{%
\includegraphics{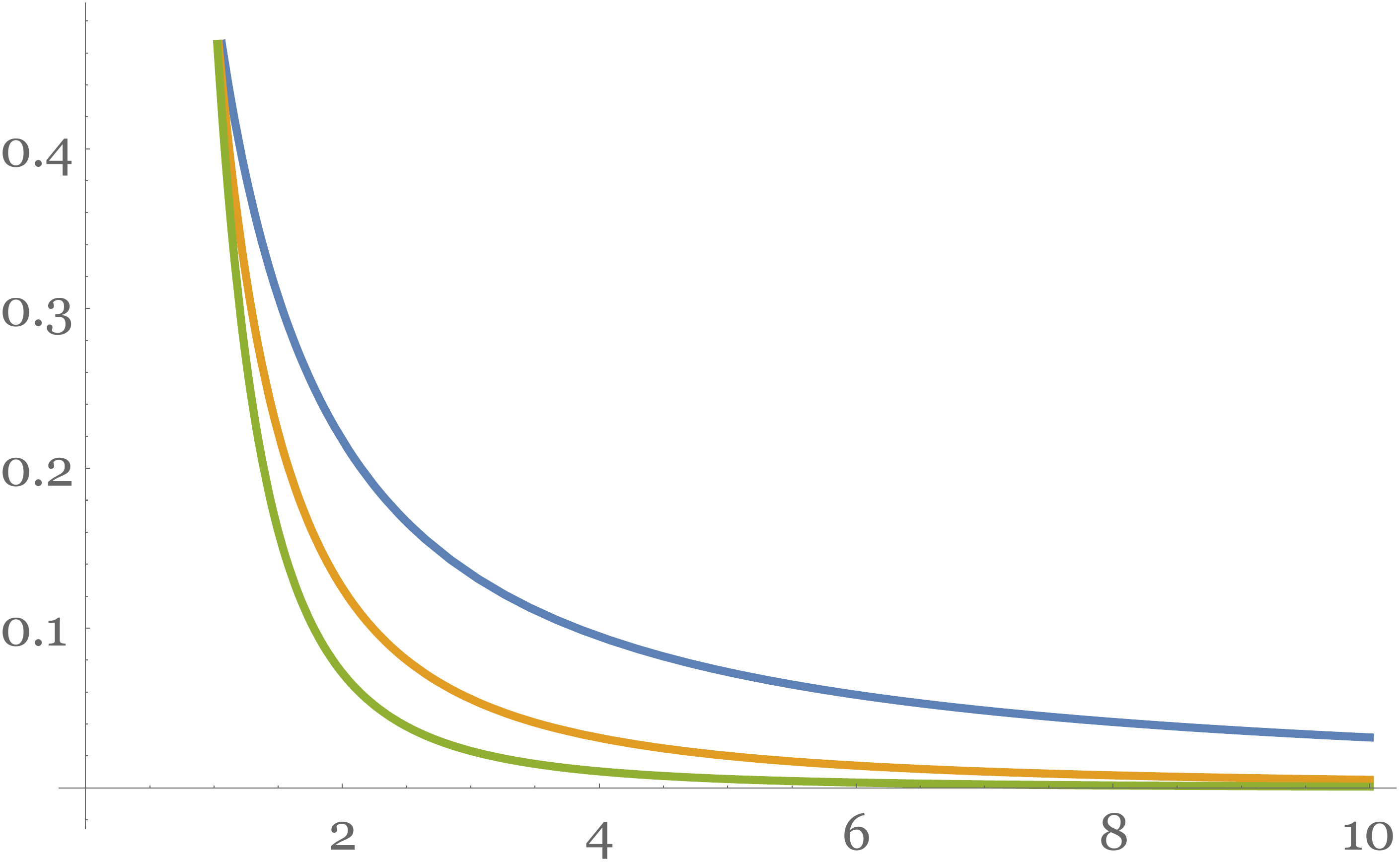}}\vskip-4mm
\caption{\small Potential energy $U(\rho)=\frac{\gamma^2}{2 \rho^{4z-2}}$ for $\gamma=1$, $z=0.8$ (top), $z=1$ (middle), $z=1.2$ (bottom).}
\label{fig3}
\end{center}
\end{figure}

It is worth mentioning that the action functional associated with Eq. (\ref{LM})
\be
S=\frac 12 \int dt \left(\dot\rho^2-\frac{\gamma^2}{\rho^{4z-2}} \right)
\ee
scales as $S'=e^{\beta(1-z)} S$ under the dilatation transformation in (\ref{TR}). In particular, for $z=1$ one can construct an extra integral of motion by applying Noether's theorem \cite{DFF}
\be\label{invD}
D=\frac 12 \rho \dot\rho-E t,
\ee
which jointly with $E$ can be used to build the general solution to (\ref{CM}) by purely algebraic means
\be\label{GS1}
\rho=\sqrt{\frac{\gamma^2+4 {\left( D+E t\right)}^2}{2 E}}.
\ee
This solution also illustrates the fact that a constant of the motion associated with the special conformal transformation is redundant for describing the conventional conformal mechanics \cite{DFF}.

As far as the single commutator (\ref{HDz}) is concerned, the parameter $z$ can be removed from the consideration by rescaling $D$. In particular, redefining the field and the temporal variable in accord with
\be\label{CT1}
\rho(t)={\tilde\rho(\tau)}^{\frac{1}{2 z-1}}, \qquad dt= \frac{ d\tau}{(2 z-1) {\tilde\rho(\tau)}^{\frac{2 z-2}{2 z-1}} },
\ee
or, equivalently
\be\label{CT}
\tilde\rho(\tau)={\rho(t)}^{2z-1}, \qquad d\tau=(2z-1){\rho(t)}^{2(z-1)} dt,
\ee
one can link the system involving arbitrary $z$, which is described by $\rho(t)$, to the model with $z=1$ featured by $\tilde\rho(\tau)$. Yet, if one is interested in the full Lifshitz algebra (see Sect. 4), the redefinition of $D$ would result in $z$ resurfacing in other commutators. In what follows, we stick to the conventional notation which keeps $z$ manifest in (\ref{HDz}).

In concluding this section, we note that given $\tilde\rho (\tau)=\pm \sqrt{\frac{\gamma^2+4 {\left( D+E \tau \right)}^2}{2 E}}$ the integral
\be
t=\int \frac{ d\tau}{(2 z-1) {\tilde\rho(\tau)}^{\frac{2 z-2}{2 z-1}} }
\nonumber
\ee
yields an expression which involves the hypergeometric function ${}_2 {F}_1 \left(\frac 12 , \frac{z-1}{2z-1},\frac 32, -\frac{4 {\left(D + E \tau\right)}^2}{\gamma^2} \right)$. Hence, it appears problematic to express $\tau$ in terms of $t$ and subsequently use the invariant $\frac 12 \tilde\rho \dot{\tilde\rho}-E \tau$ for constructing the general solution to (\ref{LM}) by purely algebraic means.

\vspace{0.5cm}

\noindent
{\bf 3. The conformal mode in a harmonic trap}\\

\noindent
The $so(2,1)$ invariance of (\ref{CM}) is preserved (see Appendix A) if one introduces into the consideration a harmonic trap potential
\be\label{HT}
\ddot\rho+\omega^2 \rho=\frac{\gamma^2}{\rho^3},
\ee
where $\omega$ is a constant frequency. The general solution to (\ref{HT}) describes oscillations around the equilibrium point $\rho_0=\sqrt{\frac{\gamma}{\omega}}$
\be\label{GS2}
\rho(t)=\frac{1}{\omega} \sqrt{E-\sqrt{E^2-\omega^2 \gamma^2} \cos{\left(2\omega(t+t_0) \right)}},
\ee
where $E$, $t_0$ are constants of integration, and it reduces to (\ref{GS1}) in the limit $\omega \to 0$.

Redefining the evolution parameter and the field in accord with (\ref{CT}), one obtains an analogue of (\ref{HT}) involving an arbitrary dynamical exponent $z$
\be\label{HT1}
\ddot\rho+(2z-1)\omega^2 \rho^{4z-3}=\frac{(2z-1)\gamma^2}{\rho^{4z-1}}.
\ee

\begin{figure}[ht]
\begin{center}
\resizebox{0.5\textwidth}{!}{%
\includegraphics{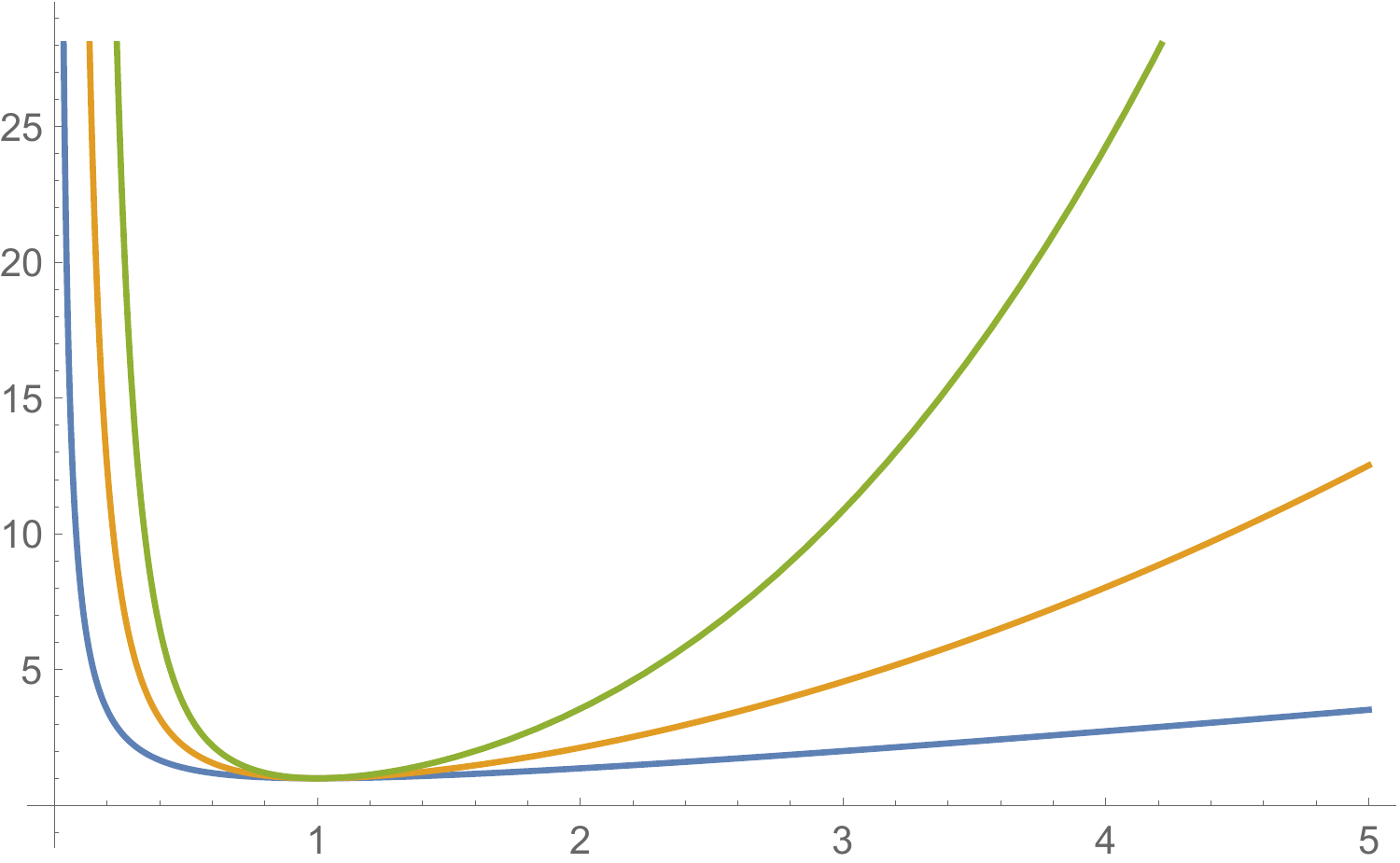}}\vskip-4mm
\caption{\small The graph of $U(\rho)=\frac 12 \left(\omega^2 \rho^{4z-2}+\frac{\gamma^2}{\rho^{4z-2}} \right),$ for $\omega=1$, $\gamma=1$, $z=0.8$ (bottom), $z=1$ (middle), $z=1.2$ (top).}
\label{fig3}
\end{center}
\end{figure}

\noindent
The latter is characterized by the conserved energy
\be\label{ConE}
E=\frac 12 \left(\dot\rho^2+\omega^2 \rho^{4z-2}+\frac{\gamma^2}{\rho^{4z-2}} \right).
\ee
Similarly to (\ref{HT}), the generalized system describes oscillations around the equilibrium point
\be
\rho_0={\left(\frac{\gamma}{\omega} \right)}^{\frac{1}{4z-2}}.
\ee
Fig. 3 plots the potential energy for $\omega=1$, $\gamma=1$, $z=0.8$ (bottom), $z=1$ (middle), $z=1.2$ (top).

Because it proves problematic to evaluate the integral
\be
t=\int \frac{ d\tau}{(2 z-1) {\left(\frac{1}{\omega} \sqrt{E-\sqrt{E^2-\omega^2 \gamma^2} \cos{\left(2\omega(\tau+\tau_0) \right)}}\right)}^{\frac{2 z-2}{2 z-1}} }
\nonumber
\ee
in a closed form, only implicit solutions to (\ref{HT1}) can be obtained via the link (\ref{CT1}) to $z=1$ variant in (\ref{HT}). For a similar reason, the conformal invariance of (\ref{HT}) is no longer transparent after switching to the partner equation (\ref{HT1}).

Concluding this section, we note that allowing $\omega$ in (\ref{HT1}) to be time--dependent, i.e. $\omega \to \Omega(t)$, one obtains a generalization of the Ermakov--Milne--Pinney equation\footnote{For a recent account of the Ermakov--Milne--Pinney equation see \cite{CGGH}.}
\be\label{GEE}
\ddot\rho+(2z-1)\Omega^2 \rho^{4z-3}=\frac{(2z-1)\gamma^2}{\rho^{4z-1}},
\ee
which involves an arbitrary dynamical exponent $z$. Although $\omega \to \Omega(t)$ breaks the Lifshitz symmetry, in Sect. 5 we construct a metric of the Lorentzian signature in $(d+2)$--dimensional spacetime and the energy--momentum tensor, which involve a cosmic scale factor $\rho(t)$ and lead to (\ref{GEE}) after imposing the Einstein equations, thus generalizing the recent studies in \cite{G,CGGH}.

\vspace{0.5cm}

\noindent
{\bf 4. Dynamical realizations of the Lifshitz group in mechanics}\\

\noindent
Let us now turn to the full Lifshitz algebra
\bea\label{LA}
&&
[H,D]={\rm i} z  H, \qquad [D,P_i]=-\frac{{\rm i}}{2} P_i, \qquad [M_{ij},P_k]=-{\rm i} \delta_{ik} P_j+{\rm i} \delta_{jk} P_i,
\nonumber\\[2pt]
&&
[M_{ij},M_{kl}]=-{\rm i} \delta_{ik} M_{jl}-{\rm i} \delta_{jl} M_{ik}+{\rm i} \delta_{il} M_{jk}+{\rm i} \delta_{jk} M_{il},
\eea
where $P_i$ and $M_{ij}$, $i=1,\dots,d$, are the generators of spatial translations and rotations, respectively, and $\delta_{ij}$ is the Kronecker delta. In a nonrelativistic spacetime parameterised by $t$ and $x_i$, $i=1,\dots,d$,  the algebra can be realized by the differential operators
\be\label{LA1}
H={\rm i} \partial_t, \qquad D={\rm i} z t \partial_t+\frac{{\rm i}}{2} x_i \partial_i, \qquad P_i={\rm i} \partial_i, \qquad  M_{ij}={\rm i}x_i \partial_j-{\rm i} x_j \partial_i,
\ee
where $\partial_t=\frac{\partial}{\partial t}$, $\partial_i=\frac{\partial}{\partial x_i}$.

Note that, because the temporal and spatial coordinates scale differently under the dilatation, a conventional kinetic term can not be used to construct invariant action functionals. A way out is to introduce an extra coordinate, transforming similarly to the spatial coordinates, and use it for building the Lifshitz--invariant derivative and field combinations. The method of nonlinear realizations \cite{CWZ} provides a rigorous way of implementing the idea. A similar consideration of the $\ell$--conformal Galilei group has been reported in \cite{FIL,GM}.

In order to construct dynamical systems invariant under transformations forming the Lifshitz group, one starts with the coset space element
\be\label{CSE}
g=e^{{\rm i}tH} e^{{\rm i} u(t) D} e^{{\rm i} x_i(t) P_i} \times {\mbox SO(d)} ,
\ee
and then computes $g^{-1} d g$, which gives rise to the Maurer--Cartan one--forms
\be\label{MCI1}
e^{-z u} dt, \qquad  du, \qquad dx_i+\frac 12 x_i du.
\ee
The forms hold invariant under the Lifshitz transformations acting on the temporal variable $t$ and the Goldstone fields $u(t)$, $x_i(t)$
\begin{align}\label{LTR}
&
t'=t+\alpha, && u'(t')=u(t), && x'_i(t')=x_i (t);
\nonumber\\[2pt]
&
t'=e^{\beta z} t, && u'(t')=u(t)+\beta, && x'_i(t')=x_i (t);
\nonumber\\[2pt]
&
t'=t, && u'(t')=u(t), && x'_i(t')=x_i (t)+a_i e^{-\frac{u}{2}},
\end{align}
which can be obtained by analyzing the left action of the group on the coset space $g'=e^{{\rm i}\alpha H} e^{{\rm i} \beta D} e^{{\rm i} a_i(t) P_i} \cdot g$.

Note that a parametrization of the coset space element chosen in (\ref{CSE}) results in the dilatation transformation which leaves $x_i(t)$ inert. Taking into account the identity
\be
e^{{\rm i}tH} e^{{\rm i} u(t) D} e^{{\rm i} x_i(t) P_i}=e^{{\rm i} {\tilde x}_i (t) P_i} e^{{\rm i}tH} e^{{\rm i} u(t) D}, \qquad {\tilde x}_i (t)=e^{\frac{u(t)}{2}} x_i (t),
\ee
which is readily established by making use of the Baker--Campbell--Hausdorff formula (\ref{ser}),
one can verify that the pair $(t,{\tilde x}_i)$ transforms in the conventional way under the temporal translation, dilatation and spatial translation
\begin{align}\label{LTR1}
&
t'=t+\alpha, && \rho'(t')=\rho(t), && {\tilde x}'_i(t')={\tilde x}_i (t);
\nonumber\\[2pt]
&
t'=e^{\beta z} t,  && \rho'(t')=e^{\frac{\beta}{2}} \rho(t), && {\tilde x}'_i(t')=e^{\frac{\beta}{2}}{\tilde x}_i (t);
\nonumber\\[2pt]
&
t'=t, && \rho'(t')=\rho(t), && {\tilde x}'_i(t')={\tilde x}_i (t)+a_i,
\end{align}
where we switched form $u$ to $\rho=e^{\frac{u}{2}}$.

From (\ref{MCI1}) one gets the invariant derivative and fields
\be\label{invar}
\mathcal{D}=e^{z u} \frac{d}{dt}, \qquad  \mathcal{D} u, \qquad \mathcal{D} x_i+\frac 12 x_i \mathcal{D} u,
\ee
which are the building blocks for constructing equations of motion. For the conformal mode $\rho=e^{\frac{u}{2}}$ it seems reasonable to accept the variant (\ref{LM}) in Sect. 2, while one is at liberty to choose any combination for the spatial coordinates, including higher derivative variants.

For example, the equation
\be\label{OSC}
\mathcal{D} \left(\mathcal{D} x_i+\frac 12 x_i \mathcal{D} u \right)+w \mathcal{D} u  \left(\mathcal{D} x_i+\frac 12 x_i \mathcal{D} u \right)=0,
\ee
where $w$ is a constant, describes the oscillator
\be\label{LO}
{\ddot x}_i+\gamma(t) {\dot x}_i+\omega(t) x_i=0,
\ee
involving the time--dependent frequency and damping coefficients
\be
\gamma (t)=(2z+2w+1) \frac{\dot\rho}{\rho} , \qquad \omega(t)=\frac{\ddot\rho}{\rho}+(2z+2w-1){\left(\frac{\dot\rho}{\rho}\right)}^2.
\ee
Interestingly enough, for $w=\frac{1-2z}{2}$ it can be cast into the total derivative form
\be\label{TDer}
{\left( \rho x_i \right)}^{\ddot{}}=0,
\ee
which holds invariant under a larger symmetry group (see Appendix B). In particular, choosing $\rho$ to be the evolution parameter, from (\ref{GS}) one gets
\be\label{PSol}
x_i (\rho)=\frac{\alpha_i}{\rho}+ {}_2 F_1 \left(\frac 12,\frac{1}{2-4z};1+\frac{1}{2-4z};\frac{\gamma^2}{2 E \rho^{4z-2}} \right) \beta_i ,
\ee
where $\alpha_i$, $\beta_i$ are constant vectors and $\rho \in (\rho_0 (z),\infty)$, where $\rho_0 (z)$ is the turning point (\ref{TP}). Being the sum of two vectors with varying length, (\ref{PSol}) describes a curve on a two--dimensional plane in $d$ dimensions. Fig. 4 displays the parametric plot $(x_1(\rho),x_2(\rho))$ for $E=\frac 12$, $\gamma=1$, $\alpha_i=(1,1)$, $\beta_i=(1,-1)$, $z=0.8$ (left), $z=1$ (middle), $z=1.2$ (right), $\rho \in [1,8]$.

\begin{figure}[ht]
\begin{center}
\resizebox{0.4\textwidth}{!}{%
\includegraphics{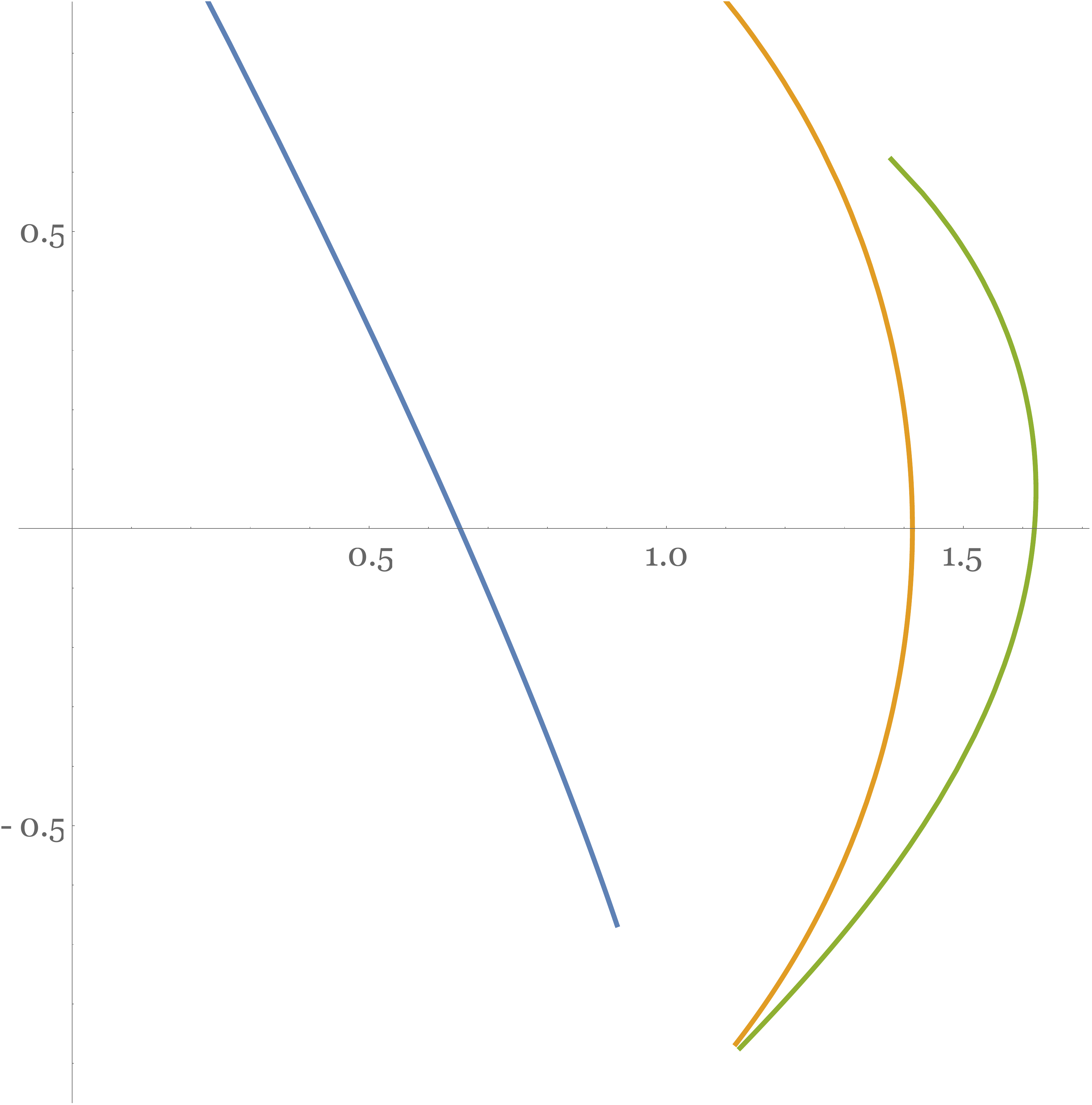}}\vskip-4mm
\caption{\small The parametric plot $(x_1(\rho),x_2(\rho))$ for $E=\frac 12$, $\gamma=1$, $\alpha_i=(1,1)$, $\beta_i=(1,-1)$, $z=0.8$ (left), $z=1$ (middle), $z=1.2$ (right), $\rho \in [1,8]$.}
\label{fig3}
\end{center}
\end{figure}

Interesting higher derivative models invariant under the Lifshitz group can be constructed by making recourse to curvature invariants of a curve in arbitrary dimension (see e.g. \cite{Am}). For example, a regular curve in three--dimensional space $x_i=x_i (t)$, $i=1,2,3$, is characterized by the curvature and torsion
\be
\kappa_1=\frac{|\dot{\vec{x}} \times \ddot{\vec{x}} |}{{|\dot{\vec{x}}|}^3}, \qquad \kappa_2=\frac{\left(\dot{\vec{x}} \times \ddot{\vec{x}}\right)\cdot \dddot{\vec{x}}}{{|\dot{\vec{x}} \times \ddot{\vec{x}} |}^2}.
\ee
Implementing the substitution
\be
{\dot x}_i  \to  \mathcal{D} x_i+\frac 12 x_i \mathcal{D} u, \qquad {\ddot x}_i \to   \mathcal{D} \left(\mathcal{D} x_i+\frac 12 x_i \mathcal{D} u \right), \qquad {\dddot x}_i \to   \mathcal{D}^2 \left(\mathcal{D} x_i+\frac 12 x_i \mathcal{D} u \right),
\ee
in $\kappa_1$ and $\kappa_2$ above and using them to construct a Lagrangian density $\mathcal{L}(\kappa_1,\kappa_2)$ (in doing so $\rho=e^{\frac{u}{2}}$ is to be regarded as a fixed function obeying (\ref{LM})), one can build a geometrically inspired action functional
\be\label{act}
S=\int  dt e^{-z u} \mathcal{L}(\kappa_1,\kappa_2),
\ee
from which an equation of motion for $x_i$ can be obtained. The model (\ref{act}) will be studied in more details elsewhere.

\vspace{0.5cm}

\noindent
{\bf 5. Conformal mode as a cosmic scale factor}\\

\noindent
Let us turn back to the conformal mechanics in the harmonic trap studied in Sect. 3, choose $z=1$ and rewrite the conserved energy (\ref{ConE}) as
\be\label{ConE1}
\dot\rho^2=-\frac{\gamma^2}{\rho^{2}}-\omega^2 \rho^{2}+2E.
\ee
Remarkably enough, this equation is akin to the Friedmann equation describing the radiation dominated Universe
\be\label{FE}
\dot\rho^2=\frac{C}{\rho^{2}}+\frac{\Lambda}{3}\rho^{2}-\kappa,
\ee
where $\rho(t)$ is a cosmic scale factor in the Friedmann--Robertson--Walker metric, $\Lambda$ is a cosmological constant, $\kappa=-1,0,1$, and $C$ is a positive constant entering the equation of state (see e.g. \cite{RI}). If $\gamma^2$ in (\ref{ConE1}) were negative, the equation would be a particular instance of (\ref{FE}), and $\rho(t)$ in (\ref{ConE1}) might have been interpreted as a cosmic scale factor relevant for describing the radiation dominated Universe. Unfortunately, changing $\gamma^2 \to -\gamma^2$ in (\ref{ConE1}) makes the original conformal mechanics unstable.

Although, (\ref{ConE1}) is not directly applicable to realistic cosmology, in this section we discuss Lorentzian metrics, for which the conformal mode $\rho(t)$ in Sect. 3 represents a cosmic scale factor. Maintaining the Lifshitz isometry group for a curved metric turns out to be problematic and the formalism is more suited for dealing with the generalized Ermakov--Milne--Pinney equation (\ref{GEE}).

An elegant geometric reformulation of a classical mechanics model with $d$ degrees of freedom $x_i$, $i=1,\dots,d$, and potential energy $U(t,x)$ was achieved in \cite{E} (see also \cite{DBKP,DGH}) in terms of null geodesics associated with the $(d+2)$--dimensional Lorentzian metric
\be\label{EisM}
ds^2=-2 U({t},x) d t^2-dt dv+dx_i dx_i,
\ee
where $t$ is a temporal variable and $v$ is an extra coordinate giving rise to the covariantly constant null Killing vector field
\be\label{Xi}
\xi^\mu \partial_\mu=\partial_v,
\ee
with $\partial_\mu=\frac{\partial}{\partial y^\mu}$ and $y^\mu=(t,v,x_i)$. The latter implies
that (\ref{EisM}) belongs to the Kundt class. In this section, we discuss an extension of (\ref{EisM}) which links to the generalized Ermakov--Milne--Pinney equation (\ref{GEE}).

Following \cite{G,CGGH}, let us focus on the oscillator potential $U(x)=\gamma^2 x_i x_i$ and introduce into the consideration a cosmic scale factor $\rho(t)$
\be
ds^2=-\frac{(2z-1)\gamma^2 x_i x_i}{\rho^{4z-2}} dt^2-dt dv+\rho^2 dx_i dx_i.
\ee
Inverting the metric and computing the Christoffel symbols
\begin{align}
&
\Gamma^v_{tt}=-\frac{2{(2z-1)}^2 \gamma^2 x_i x_i \dot\rho}{\rho^{4z-1}}, && \Gamma^v_{t i}=\frac{2(2z-1)\gamma^2 x_i}{\rho^{4z-2}}, && \Gamma^v_{ij}=2 \rho \dot\rho \delta_{ij},
\nonumber\\[2pt]
&
\Gamma^i_{tt}=\frac{(2z-1)\gamma^2 x_i}{\rho^{4z}}, && \Gamma^i_{tj}=\frac{\dot\rho}{\rho} \delta_{ij},
\end{align}
where we have split the index $\mu=(t,v,i)$, $i=1,\dots,d$, and denoted $\dot\rho=\frac{d\rho}{dt}$, one can verify that the temporal variable $t$ is related to the affine parameter $\lambda$ along a null geodesic
\be
t=t_1 \lambda+t_0,
\ee
where $t_0$ and $t_1$ are constants of integration, the equation of motion for $x_i$ reads
\be\label{X}
{\left( \rho x_i \right)}^{\ddot{}}-\left(\frac{\ddot\rho}{\rho}-\frac{(2z-1)\gamma^2}{\rho^{4z}} \right) \rho x_i=0,
\ee
where the dot designates the derivative with respect to $t$,
while the evolution of $v(t)$ over time is governed by the null geodesic equation $ds^2=0$
\be
\dot v= -\frac{(2z-1)\gamma^2 x_i x_i}{\rho^{4z-2}}+\rho^2 {\dot x}_i {\dot x}_i.
\ee
In particular, if $\rho$ is chosen to obey (\ref{LM}), Eq. (\ref{X}) reduces to (\ref{TDer}).

The corresponding geometry, however, is flat as can be seen by implementing the coordinate transformation
\be
t'=t, \qquad x'_i=\rho x_i, \qquad v'=v+\rho \dot\rho x_i x_i,
\ee
which brings the metric to the form
\be\label{M1}
ds^2=\left(\frac{\ddot\rho}{\rho}-\frac{(2z-1) \gamma^2}{\rho^{4z}}\right) x'_i x'_i dt'^2-dt' dv'+dx'_i dx'_i.
\ee
If $\rho$ obeys (\ref{LM}), the first term in  (\ref{M1}) vanishes and $ds^2$ simplifies to the $5d$ Minkowski metric, $(t',v')$ being the double--null coordinates.

The picture becomes more interesting if, by analogy with \cite{CGGH}, one decides to impose the generalized Ermakov--Milne--Pinney equation (\ref{GEE}) on the cosmic scale factor $\rho$.
In this case the metric (\ref{M1}) is no longer flat and
Eq. (\ref{GEE}) can be derived from the Einstein equations
\be
R_{\mu\nu}=8\pi T_{\mu\nu}
\ee
by introducing the energy--momentum tensor $T_{\mu\nu}$
\be\label{EMT}
T_{\mu\nu}=\frac{(2z-1)d}{2\pi} \Omega^2 \rho^{4(z-1)} \xi_\mu \xi_\nu, \qquad {T^\mu}_\mu=0, \qquad
\nabla^\mu T_{\mu\nu}=0,
\ee
where $\xi_\mu$ is the covariantly constant null Killing vector field (\ref{Xi}). In particular, Eq. (\ref{X}) turns into
\be
{\ddot x'}_i +(2z-1)\Omega^2 \rho^{4(z-1)} x'_i =0,
\ee
where $x'_i=\rho x_i$, which reduces to the time--dependent harmonic oscillator equation for $z=1$.

\vspace{0.5cm}

\noindent
{\bf 6. Lorentzian metrics with the Lifshitz isometry group}\\

\noindent
The method of nonlinear realizations can also be used for building Lorentzian metrics with the Lifshitz isometry group, which are relevant for describing gravity duals to field theories enjoying the Lifshitz symmetry \cite{MT,KLM}. For example, focusing on the group manifold
\be\label{CSE3}
g=e^{{\rm i}tH} e^{{\rm i} u D} e^{{\rm i} X_i P_i} \times {\mbox SO(d)},
\ee
which is parameterized by the coordinates $(t,u,X_i)$, $i=1,\dots,d$,
and computing the Maurer--Cartan invariants as above
\be\label{MCI3}
e^{-z u} dt, \qquad  du, \qquad dX_i+\frac 12 X_i du,
\ee
one gets the natural building blocks for constructing invariant quadratic forms (metrics). In particular, redefining the coordinates
$r=e^{-\frac{u}{2}}$, $x_i=e^{\frac{u}{2}} X_i$, which bring (\ref{MCI3}) to the form familiar from studying near horizon black hole geometries
\be\label{MCI4}
r^{2z} dt, \qquad  \frac{dr}{r}, \qquad r dx_i,
\ee
one obtains the metric\footnote{The Lifshitz algebra (\ref{LA}) reproduces that in \cite{KLM} after rescaling $2D\to D$, $2z \to z$.}
\be\label{LMetr}
ds^2=-r^{4z} dt^2+\frac{dr^2}{r^2}+r^2 dx_i dx_i,
\ee
which is the key ingredient in describing the Lifshitz holography \cite{KLM}. Note that (\ref{LMetr}) reduces to the ${AdS_{d+2}}$ metric in the Poincar\'e coordinates for $2z=1$.

Extending the Lifshitz algebra by the Galilei boost generator $K_i$ (see Appendix B), similarly enlarging the group element (\ref{CSE3}), and repeating the steps above,
one derives the invariants
\be
r^{2z} dt, \qquad  \frac{dr}{r}, \qquad r (dx_i-y_i dt), \qquad \frac{d y_i}{r^{2z-1}},
\ee
where $y_i$ are extra coordinates associated with $K_i$, which give rise to a natural generalization of (\ref{LMetr})
\be\label{LMetr1}
ds^2=-r^{4z} dt^2+\frac{dr^2}{r^2}+r^2 (dx_i-y_i dt)(dx_i-y_i dt)+\frac{\mu^2}{r^{4z-2}} dy_i dy_i+\frac{\nu^2}{r^{2z-2}} dy_i (dx_i-y_i dt),
\ee
$\mu$, $\nu$ being constant parameters obeying $\mu^2>\nu^4$. The quadratic form (\ref{LMetr1}) holds invariant under temporal translation and spatial rotation, as well as under the anisotropic conformal transformation and the Galilei boost
\begin{align}
&
t'=\lambda^{2z} t, && r'=\lambda^{-1} r, && x'_i=\lambda x_i, && y'_i=\lambda^{1-2z} y_i;
\nonumber\\[2pt]
&
x'_i=x_i+v_i t, && y'_i=y_i+v_i, && &&
\end{align}
where $\lambda$ and $v_i$ are finite transformation parameters. It would be interesting to study whether (\ref{LMetr1}) can be obtained as a solution to field equations of gravity coupled to some matter content and whether it can be used for holographic applications in the spirit of \cite{KLM}.

\vspace{0.5cm}

\noindent
{\bf 7. Conclusion}\\

\noindent
To summarize, in this work dynamical realizations of the Lifshitz group were studied. A generalization of the $1d$ conformal mechanics \cite{DFF} was constructed, which involved an arbitrary dynamical exponent $z$. A similar generalization of the Ermakov--Milne--Pinney equation was proposed. The method of nonlinear realizations \cite{CWZ} was used to determine the Lifshitz--invariant derivative and field combinations, which enabled us to construct dynamical systems enjoying such symmetry. Extending the recent studies in \cite{G,CGGH}, a metric of the Lorentzian signature in $(d+2)$--dimensional spacetime and the energy--momentum tensor were constructed, which led to the generalized Ermakov--Milne--Pinney equation upon imposing the Einstein equations.
The corresponding null geodesic equations were shown to describe a variant of the Lifshitz oscillator driven by the conformal mode. It was demonstrated that the group--theoretic framework \cite{CWZ} could also be used for building Lorentzian metrics with the Lifshitz isometry group. In particular, a $(2d+2)$--dimensional extension of the $(d+2)$--dimensional metric in \cite{KLM} was constructed, which enjoyed an extra invariance under the Galilei boost.

Turning to possible further developments, it would be interesting to analyze in more detail the issue of integrability for the models in Sect. 4. Explicit solutions to the generalized Ermakov--Milne--Pinney equation are worth studying as well. It is interesting to explore whether the metric (\ref{LMetr1}) in Sect. 6 can be obtained as a solution to field equations of gravity coupled to some matter content and whether it can be used for holographic applications in the spirit of \cite{KLM}.

\vspace{0.5cm}

\noindent{\bf Acknowledgements}\\

\noindent
This work is supported by the Russian Foundation for Basic Research, grant No 20-52-12003.

\vspace{0.5cm}

\noindent
{\bf Appendix A: Symmetries of Eq. (\ref{HT})}\\

\noindent
In this Appendix, we discuss symmetries of Eq. (\ref{HT}). Demanding (\ref{HT}) to hold invariant under the transformation $t'=a(t)$, $\rho'(t')=b(t) \rho(t)$,\footnote{One could try a more general ansatz $t'=a(t)$, $\rho'(t')=b(t) \varphi(\rho(t))$, $\varphi(\rho(t))$ being an arbitrary function. Yet, the requirement that (\ref{HT}) be invariant under the transformation results in $\frac{d^2 \varphi}{d \rho^2}=0$. } one gets two differential equations
\be
b \ddot b -2 {\dot b}^2+\omega^2 b^2 \left( b^4-1\right)=0, \qquad \dot a=b^2.
\nonumber
\ee
The general solution to these equations involves three constants of integration, say $\alpha$, $\beta$, $\sigma$, which give rise to the following symmetry transformations of (\ref{HT})
\bea
&&
t'=t+\alpha, \qquad \rho'(t')=\rho(t);
\nonumber\\[2pt]
&&
t'=\frac{1}{\omega} \arctan{\left(\left(\sqrt{1+\beta^2}-\beta \right) \tan{(\omega t)}\right)}, \quad \rho'(t')=\frac{\rho(t)}{\sqrt{\sqrt{1+\beta^2}+\beta \cos{(2\omega t)}}};
\nonumber\\[2pt]
&&
t'=\frac{1}{\omega} \arctan{\left(\sigma + \sqrt{1+\sigma^2} \tan{(\omega t)}\right)}, \quad
\rho'(t')=\frac{\rho(t)}{\sqrt{\sqrt{1+\sigma^2}+\sigma \sin{(2\omega t)}}}.
\nonumber
\eea
Expanding each function of $\beta$ and $\sigma$ into the Taylor series up to the first order, one obtains the generators of infinitesimal transformations
\bea
&&
H={\rm i} \partial_t -\omega^2 C, \quad
D=\frac{{\rm i}}{2\omega} \sin{(2\omega t)} \partial_t+\frac{{\rm i}}{2} \cos{(2 \omega t)} \rho \partial_\rho,
\nonumber\\[2pt]
&&
C=\frac{{\rm i}}{2 \omega^2} \left(1-\cos{(2 \omega t)} \right) \partial_t+\frac{{\rm i}}{2 \omega} \sin{(2\omega t)} \rho \partial_\rho,
\nonumber
\eea
which obey $so(2,1)$ algebra
\be
[H,D]={\rm i} H, \qquad [H,C]=2{\rm i} D, \qquad [D,C]={\rm i} C.
\nonumber
\ee
Note that in the limit $\omega \to 0$ the generators reproduce the conventional realization of $so(2,1)$
\be
H={\rm i} \partial_t, \qquad D={\rm i} t \partial_t+\frac{{\rm i}}{2} \rho \partial_\rho, \qquad C={\rm i} t^2 \partial_t+{\rm i} t \rho \partial_\rho,
\nonumber
\ee
while the reduced equations $b \ddot b -2 {\dot b}^2=0$, $\dot a=b^2$ yield
\be
t'=\frac{\alpha t+\beta}{\gamma t+\delta}, \qquad \rho'(t')={\left( \frac{dt'}{dt} \right)}^{\frac 12} \rho(t),
\ee
where $\alpha$, $\beta$, $\gamma$, $\delta$ are real constants obeying $\alpha\delta-\gamma\beta=1$. The latter point is in agreement with the analysis in \cite{DFF}.

\vspace{0.5cm}

\noindent
{\bf Appendix B: Adding the Galilei boost to the Lifshitz algebra}\\

\noindent
The Lifshitz algebra (\ref{LA}) can be extended to include the generator of Galilei boost $K_i$, which obeys the structure relations
\be
[H,K_i]={\rm i} P_i, \qquad [D,K_i]={\rm i} \left(z-\frac 12 \right) K_i, \qquad [M_{ij},K_p]=-{\rm i} \delta_{ip} K_j+{\rm i} \delta_{jp} K_i.
\nonumber
\ee
The latter adds $K_i={\rm i} t \partial_i$ to the differential operators in (\ref{LA1}). Extending the coset space element (\ref{CSE}) in a natural way
\be\label{CSE1}
g=e^{{\rm i}tH} e^{{\rm i} u(t) D} e^{{\rm i} x_i(t) P_i} e^{{\rm i} y_i(t) K_i}  \times {\mbox SO(d)},
\nonumber
\ee
and can compute the Maurer--Cartan invariants
\be
g^{-1} dg={\rm i} \omega_H H +{\rm i} \omega_D D +{\rm i} {\omega_P}_i P_i +{\rm i} {\omega_K}_i  K_i,
\nonumber
\ee
where
\be
\omega_H=e^{-z u} dt, \quad \omega_D=du, \quad {\omega_P}_i=dx_i+\frac 12 x_i du-dt e^{-z u} y_i, \quad {\omega_K}_i=d y_i-\left(z-\frac 12 \right) y_i du.
\nonumber
\ee
Setting the constraint ${\omega_P}_i/\omega_H=0$, one can express $y_i$ in terms of the other fields and their invariant derivatives
\be
y_i=\mathcal{D} x_i+\frac 12 x_i \mathcal{D} u, \qquad \mathcal{D}=e^{z u} \frac{d}{dt},
\ee
while imposing the equation of motion ${\omega_K}_i/\omega_H=0$ one reproduces a variant of the Lifshitz mechanics (\ref{OSC}) with $w=\frac{1-2z}{2}$.

\noindent


\end{document}